\documentclass[aps,prb,twocolumn,superscriptaddress]{revtex4-1}
\usepackage{graphicx}
\usepackage{amsmath,amssymb}
\usepackage{hyperref}
\usepackage{bm}
\usepackage{float}

\begin{document}

\title{Dynamics and Control of Edge States in Laser-driven Graphene Nanoribbons}

\date{\today}
\author{M.~Puviani}
\affiliation{Dipartimento di Scienze Fisiche, Informatiche e Matematiche, Universit\`a di Modena e Reggio Emilia, Via Campi 213/A, 41125 Modena, Italy }
\author{F.~Manghi}
\affiliation{Dipartimento di Scienze Fisiche, Informatiche e Matematiche, Universit\`a di Modena e Reggio Emilia, Via Campi 213/A, 41125 Modena, Italy }
\affiliation{S3, Istituto Nanoscienze -- CNR, Modena, Italy}
\author{A.~Bertoni}
\affiliation{S3, Istituto Nanoscienze -- CNR, Modena, Italy}
\begin{abstract}
An intense laser field in the high-frequency regime drives carriers in graphene nanoribbons (GNRs) out of equilibrium and creates topologically-protected edge states.
Using Floquet theory on driven GNRs, we calculate the time evolution of local excitations of these edge states and show that they exhibit a robust dynamics also in the presence of very localized lattice defects (atomic vacancies), which is characteristic of topologically non-trivial behavior.
We show how it is possible to control them by a modulated electrostatic potential: They can be fully transmitted on the same edge, reflected on the opposite one, or can be split between the two edges, in analogy with Hall edge states, making them promising candidates for flying-qubit architectures.
\end{abstract}


\maketitle

\section{Introduction}

The dynamics of quantum systems under the influence of time-periodic modulations has recently attracted growing attention for the possibility to realize  unconventional phases of matter, including topological phases.
After theoretical predictions, \cite{Kitagawa2011,Lindner2011,Inoue2010,Oka2009}  the generation and  manipulation of topological states through the application of a time-periodic perturbation has been experimentally demonstrated in different systems such as ultra-cold gases in time-dependent optical lattices \cite{Goldman2016}, periodically driven photonic waveguide lattices \cite{Maczewsky2017,Rechtsman2013}, acoustic systems \cite{He2016} and topological insulators  under circularly polarized light \cite{W.2012}.
The presence of topologically protected edge states responsible for robust one-way edge transport is the common thread between all these diverse systems.

On the theoretical side, a complete topological characterization is provided by the emergence of non-vanishing topological invariants that ensure the presence of gapless edge states and their robustness against disorder.
Topological invariants have been defined for systems in static conditions \cite{Xu2006,Hasan2010,Grandi2015,Grandi2015NJP}  and only more recently extended to the periodically driven case \cite{Rudner2013,Foa2015,Fenner2017} where new types of edge modes have been identified which cannot be accounted for using the invariants developed for the static case. \cite{Kitagawa2012,Maczewsky2017}

The topological properties of Floquet systems have also been studied in connections with transport properties: The conductance and quantum Hall response of irradiated  graphene nanoribbons (GNRs) \cite{PhysRevLett.113.236803,PhysRevLett.113.266801,PhysRevB.89.121401} and of quantum well heterostructures \cite{PhysRevLett.115.106403} have been calculated by analytical and numerical methods showing distinctive characteristics associated with the presence of chiral edge states.
The influence of disorder \cite{PhysRevLett.113.236803,PhysRevLett.115.106403} and of coupling with phonons \cite{PhysRevB.90.195429} has been used as an hallmark of a topologically protected phase.
Different kinds of defects in 2D graphene have been shown to host Floquet bound states and their chiral nature has been identified by calculating the probability current around them \cite{PhysRevB.93.245434}.

In this article, we look for another explicit signature of the robustness of edge states in a zig-zag terminated GNR driven by a circularly polarized intense laser field.
We explore the real-time evolution of a particle initialized in one of these states and study in particular how time evolution is affected by local defects and potential barriers.
Our analysis is based on the Floquet formalism, a Bloch theory in time domain which exploits time-periodicity to solve the time-dependent Schr\"{o}dinger equation, factorizing the stroboscopic time-dependence of the quasiparticle from the intrinsic periodic one.

\section{Floquet theory and time-dependent velocity}

The Hamiltonian for a lattice driven by a time-periodic electromagnetic field can be written  using the minimal coupling  \cite{PhysRevLett.108.225303}:
\begin{align}
\hat{H}_{\vec{k}} (t) = \sum_{i,i'} \sum_{l, l'} \ J_{i l, i' l'} \ e^{i \left( \vec{k} + \vec{A} (t) \right)} 
\, e^{(\vec{R}_l + \vec{\tau}_i - \vec{R}_{l'} + \vec{\tau}_{i'})} 
\nonumber \\ \hat{c}^{\dagger}_{i,l}(t) \ \hat{c}_{i',l'}(t) \ ,
\end{align}
$i,i'$ ($l,l'$) being site (cell) indices, with $\hat{c}^{\dagger}_{i,l}(t)$, $\hat{c}_{i',l'}(t) $ the corresponding time-dependent creation and annihilation operators, and $J_{i l, i' l'}$ being the nearest-neighbor hopping term.
The vector potential associated with the electromagnetic field is $\vec{A}(t) = A_0 \left( \sin( \Omega t ) \ \hat{x} + \ \cos(\Omega t ) \ \hat{y} \right)$ for clockwise circular polarization, $\Omega = 2 \pi / \text{T}$ the frequency of the driving.
Owing to the Floquet theorem, the solution of the time-dependent Schr\"{o}dinger equation
\begin{equation} \label{ShrTD}
\hat{H} (t) \ | \Psi (t) \rangle  = i  \partial_t \ | \Psi (t) \rangle 
\end{equation}
can be written in a factorized form as
\begin{equation}\label{TDsol}
| \Psi_{\alpha} (t) \rangle = e^{- i \varepsilon_{\alpha} t} | \Phi_{\alpha}(t) \rangle \ ,
\end{equation}
with $ | \Phi_{\alpha} (t+T) \rangle = | \Phi_{\alpha}(t) \rangle $.
Defining the Floquet operator as
\begin{equation}
\hat{H}_F \equiv \hat{H} (t) - i \partial_t \ ,
\end{equation}
and substituting Eq.~(\ref{TDsol}) in Eq.~(\ref{ShrTD}), we obtain a time-independent eigenvalue problem for the Floquet states, with Floquet quasienergies $\varepsilon_{\alpha}$ constant in time:
\begin{equation} \label{EigenFloq}
\hat{H}_F \ | \Phi_{\alpha}(t) \rangle = \varepsilon_{\alpha} \ | \Phi_{\alpha}(t) \rangle \quad .
\end{equation}
Since $| \Phi_{\alpha}(t) \rangle $ is periodic in time, it can be expanded in Fourier series
\begin{equation}
| \Phi_{\alpha}(t) \rangle = \sum_{n = - \infty}^{+ \infty} e^{- i n \Omega t} \ | \phi_{\alpha,n} \rangle \quad .
\end{equation}
In practice, the Fourier expansion  is truncated to include a finite number of modes, up to a cutoff $n_{max}$. This allows one to formulate the eigenvalue  problem in Eq.~(\ref{EigenFloq}) in a standard matrix form whose eigenvalues  turn out to be replicas of the static band structure with gaps opening  at their crossing points. Due to this truncation we have access to the time evolution at stroboscopic times $t=nT/n_{max}$ only. However, by extending $n_{max}$  it is possible to verify the accuracy of the dynamics at intermediate times.

The solution of the time-dependent Schr\"{o}dinger equation for a particle in a lattice written in the local basis $\varphi$ is therefore
\begin{align}\label{SoluT}
 \Psi_{\alpha}(r,t)  =  e^{- i \varepsilon_{\alpha} t } \sum_{i=1}^{N} \sum_{l=1}^{M} \sum_{n = -n_{max}}^{+ n_{max}} \sum_{\vec{k}} c_{i,n}^{\alpha} (\vec{k})  \ e^{- i n \Omega t} \nonumber \\
e^{-i \vec{k} \cdot (\vec{R}_l + \vec{\tau}_i) } \ \varphi (\vec{r}- \vec{R}_l - \vec{\tau}_i) \ ,
\end{align}
where $N$ is the number of sites per cell, $M$ the number of cells, while $n$ are the Floquet indices.
The time-dependent expectation values of observables involve these states.\cite{Manghi2017}
\\
The time-dependent expectation value of the velocity operator $\hat{\vec{v}}(t) = - i [ \hat{\vec{r}} , \hat{H}(t) ]$ can be analyzed at fixed Floquet eigenvectors  obtaining a velocity vector field in real space 
\begin{equation}\label{velocity}
\vec{v}_{\alpha} (\vec{r}_i, t) = \langle \Psi_{\alpha} (r,t) | \hat{\vec{v}}(t)| \Psi_{\alpha} (r,t) \rangle.
\end{equation}
By using the elements of operators $\hat{H}$ and $\hat{\vec{r}}$ in the localized basis, namely
\begin{equation*}
 \langle \varphi_{i,l} | \ \hat{H}(t) \ | \varphi_{i', l'} \rangle = J_{i i', l l'} \ e^{i \vec{A}(t) \cdot \left( \vec{R}_l + \vec{\tau}_i - \vec{R}_{l'} - \vec{\tau}_{i'} \right)}
\end{equation*}
and \cite{Selloni1984}
\begin{equation*}
\langle \varphi (\vec{r}- \vec{R}_l - \vec{\tau}_i)|\vec{r}|\varphi (\vec{r}- \vec{R}_{l'} - \vec{\tau}_{i'} ) \rangle = \delta_{i i'} \delta_{l l'} (\vec{R}_l + \vec{\tau}_i) \ ,
\end{equation*}
the velocity vector field turns out to be
\begin{align}
\vec{v}_{\alpha} (\vec{r}_i, t) = \sum_{\substack{i' \\ l, l'}} \sum_{n, m} \sum_{\vec{k}} \ \left( c^{\alpha}_{i',m} (\vec{k}) \right)^* c^{\alpha}_{i,n} (\vec{k})  \ e^{-i (n-m) \Omega t} \nonumber \\
e^{- i \vec{k} \cdot \left( \vec{R}_l + \vec{\tau}_i - \vec{R}_{l'} - \vec{\tau}_{i'} \right) }  \ ( \vec{R}_l + \vec{\tau}_i - \vec{R}_{l'} - \vec{\tau}_{i'} ) \nonumber \\
e^{- i \vec{A}(t) \cdot \left( \vec{R}_l + \vec{\tau}_i - \vec{R}_{l'} - \vec{\tau}_{i'} \right) } \ J_{i l , i' l'} \ .
\end{align}

\section{Dynamics of Floquet states in real space}

The probability for an electron in a given  state to move from a site $i$ at time $t_0$ to a site $j$ at time $t$ is $| U_{j,i} (t,t_0) |^2 $, where the time-evolution operator $U_{j,i}$ can be expressed interms of Floquet components as follows \cite{Shirley1965}
\begin{align}\label{timeEVO}
U_{j,i} (t,t_0) = \sum_{\alpha, n,m} e^{- i \varepsilon_{\alpha} (t-t_0)} \ e^{- i m \Omega t} \ \langle j,m | \varphi_{\alpha,m} \rangle \nonumber \\
\langle \varphi_{\alpha,n} | i,n \rangle  \ e^{i n \Omega t_0} \quad .
\end{align}
In order to have a result directly comparable with a real experiment, it is necessary to average the transition probability over the initial time $t_0$ keeping fixed the interval $\Delta_t=(t-t_0)$:
\begin{equation}\label{probab}
P_{j \leftarrow i} (\Delta_t) = \sum_m \bigg\vert \sum_{\alpha, \vec{k}} c^{\alpha}_{j,m} (\vec{k}) \left( c^{\alpha}_{i,0} (\vec{k}) \right)^* e^{-i \varepsilon_{\alpha, \vec{k}} \Delta_t} \bigg\vert ^2 \ .
\end{equation}
This allows us to determine the electron motion in real space and time, without solving any time-integral: it is noticeable in fact that the system has no memory. Therefore, this quantity can be calculated for any $t$ as large as desired, without knowing anything regarding times before $t$.

\section{Dynamics and robustness of edge states in driven GNR}

We study an ac-driven zigzag GNR extended along the $x$ axis and we consider the case of $\Omega \gg J $. With $n_{max}=3$ we get already time steps in the attoseconds regime. We verified that the dynamics at intermediate times does not substantially deviate from what is obtained with larger cutoffs.

\begin{center}
\begin{figure}[H]
 \centering \includegraphics[width=9cm]{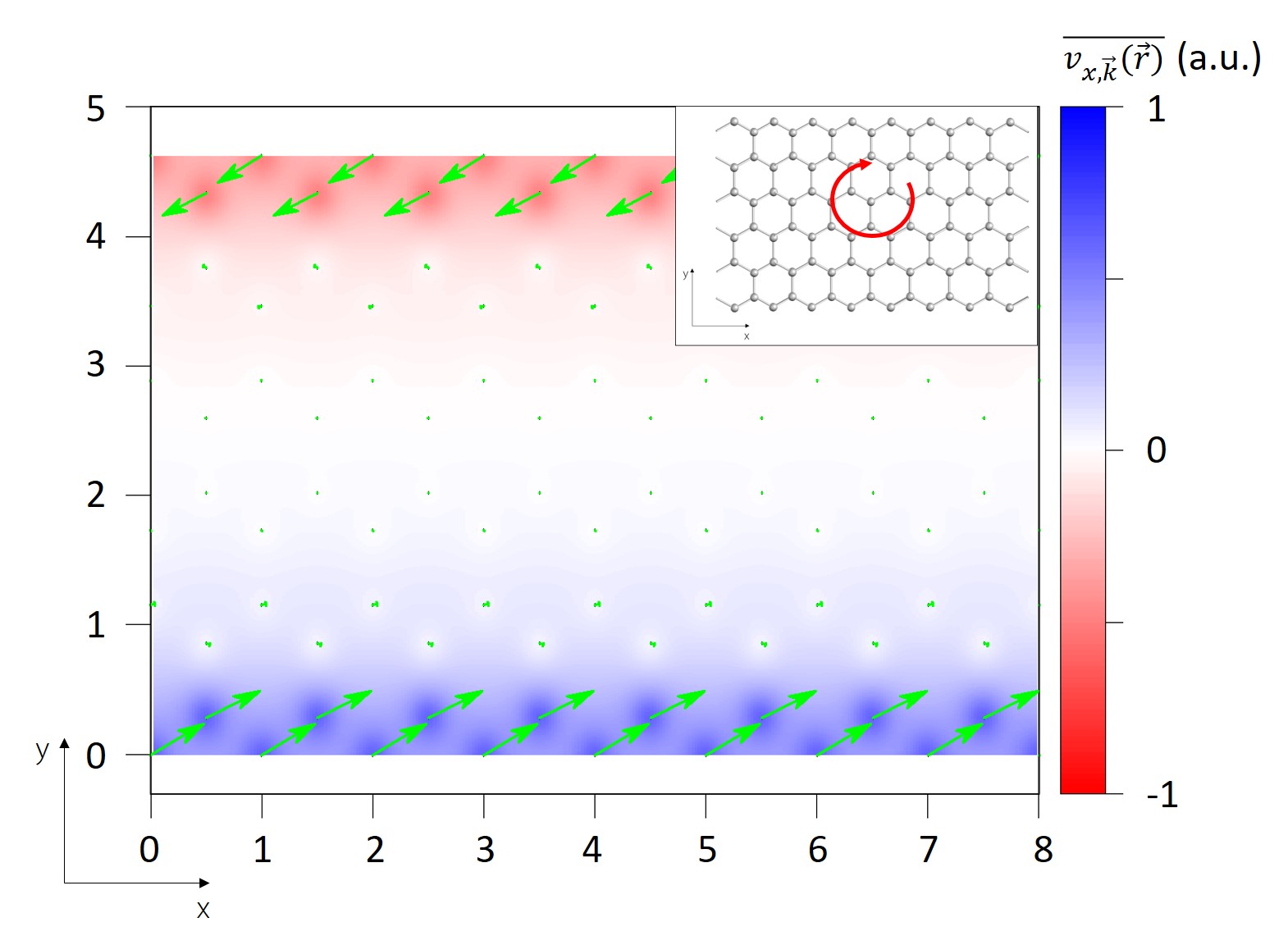}
 \caption{\label{fig1:edge_velocity} (color online) Time-averaged velocity plot in real space for the edge states of the GNR. The arrows represent the velocity averaged over a period of the external field for each site, while the color scale is associated with the projection of such vector along the $x$ direction of the nanoribbon.
The inset shows a portion of the GNR lattice, and the curved arrow indicates the circular laser polarization. }
\end{figure}
\end{center}

\begin{center}
\begin{figure}[H]
 \centering \includegraphics[width=9cm]{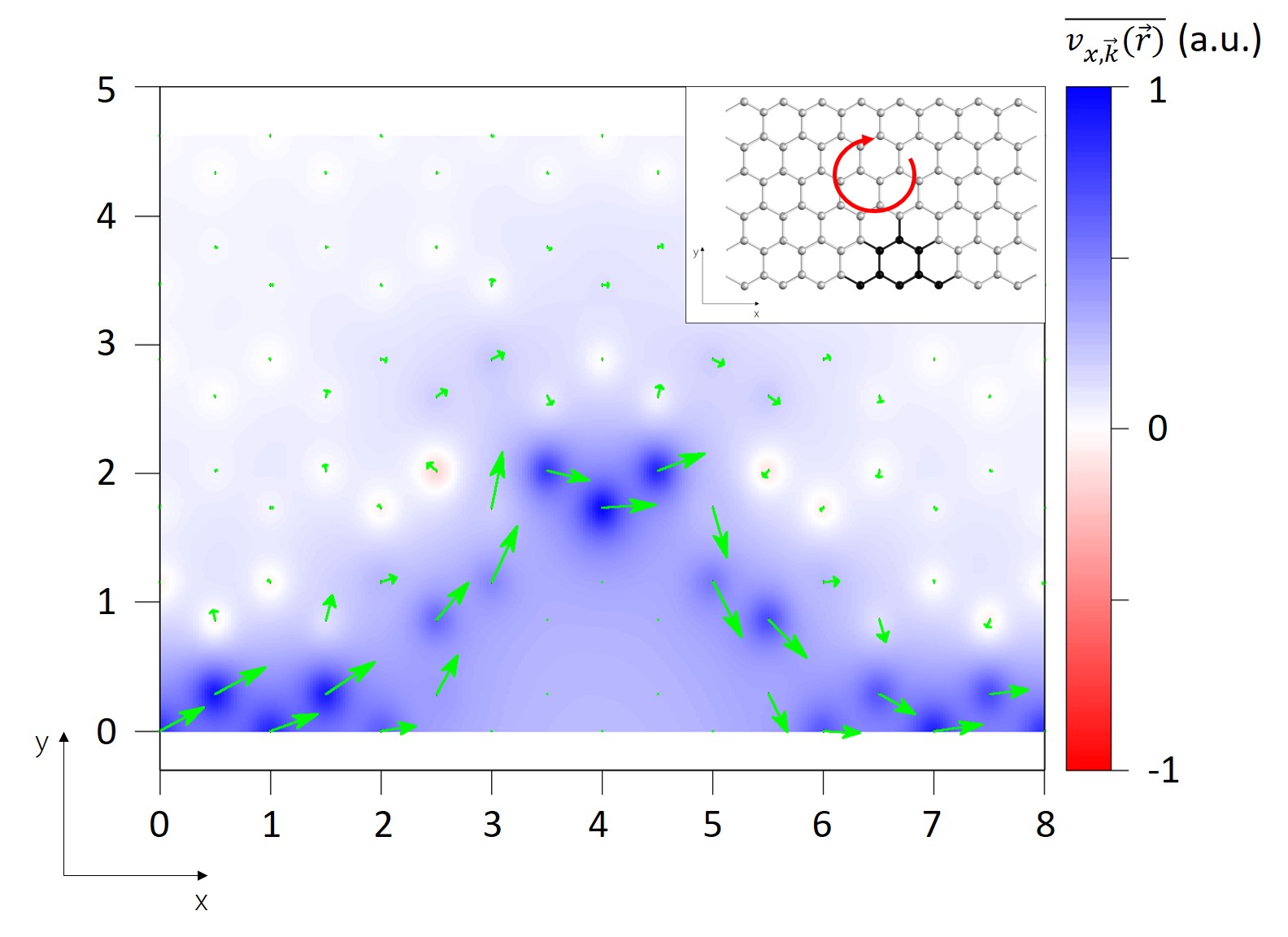}
 \caption{\label{fig2:Low1Curr} (color online) Time-averaged velocity in real space for an edge state localized in the lower part of the nanoribbon and propagating rightward. A six-atom vacancy is present at the lower edge as shown in the inset, where the bold dots indicate atoms not included in the calculation and where the curved arrow indicates the circular laser polarization.  The arrows represent the velocity averaged over a period of the external field for each site, while the color scale is associated with the projection of these vectors on the $x$ axis.}
\end{figure}
\end{center}

\begin{center}
\begin{figure}[H]
 \centering \includegraphics[width=9cm]{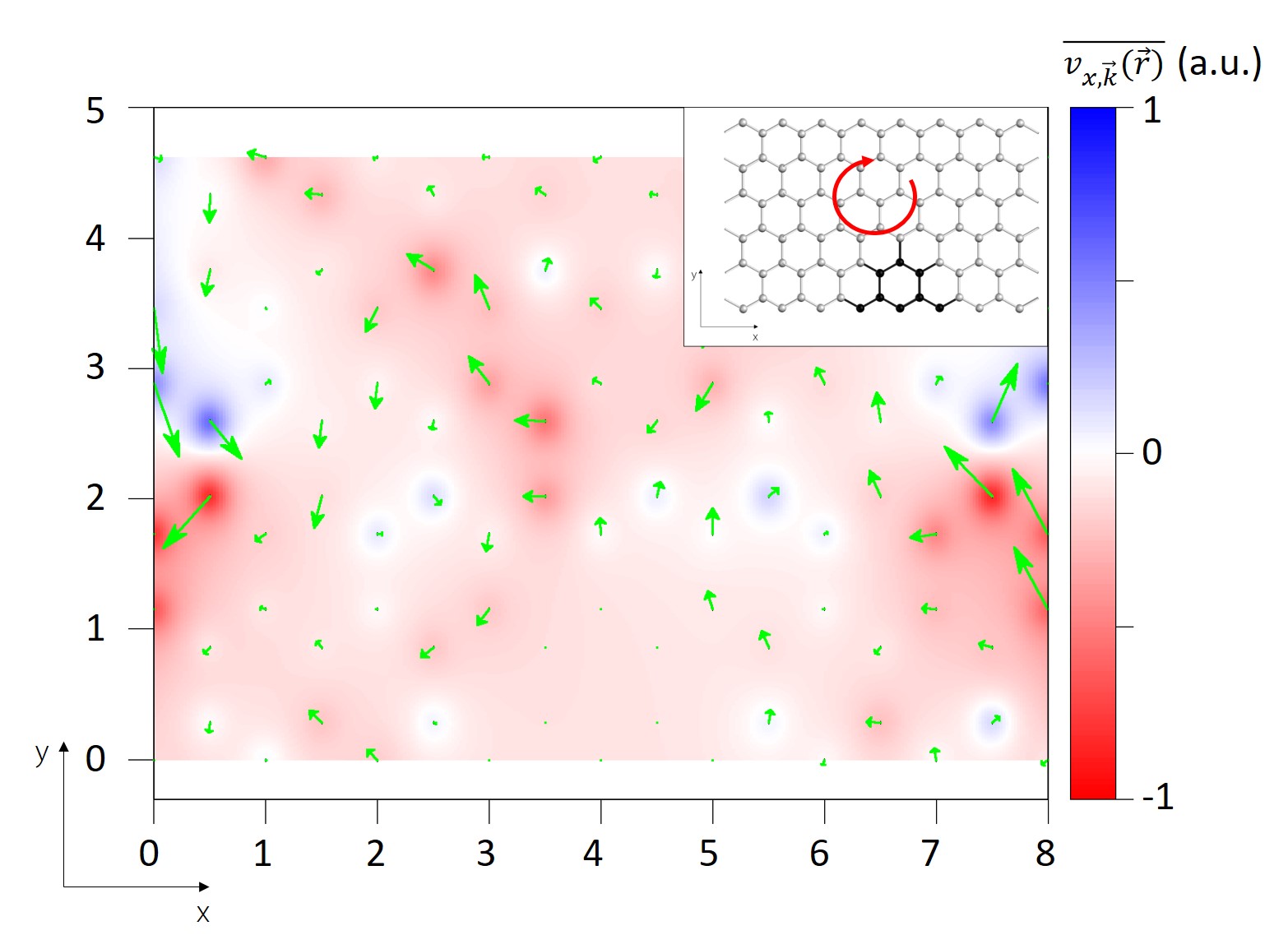}
 \caption{\label{fig3:bulkCurr} (color online) Same as Fig.~\ref{fig2:Low1Curr}, but for  a state  which is not localized at the edge. }
\end{figure}
\end{center}

In the presence of circularly-polarized light, edge states are localized either at the upper or at the lower GNR edge, and are characterized by unidirectional and opposite values of the velocity vector field. This is shown in Fig.~\ref{fig1:edge_velocity} where we report the time-averaged velocity $\overline{\vec{v}_{\alpha, \vec{k}} (\vec{r}_i)} = \dfrac{1}{T} \int_{0}^{T} \vec{v}_{\alpha, \vec{k}} (\vec{r}_i, t) \ dt $ calculated  over the two edge states.
Now, we want to include a defect at one edge. To this aim, we extend the ribbon unit cell and build a supercell of $12 \times 27$ atoms (the ribbon is 12-atoms wide), with  periodic boundary conditions along the $x$ axis. We selectively remove atoms at one edge, as shown in the inset of Fig.~\ref{fig2:Low1Curr}, creating a multi-atom vacancy.
Gapless edge states characteristic of driven GNRs \cite{Manghi2017} persist in the presence of this rather strong perturbation and, interestingly, they give rise to the peculiar time-averaged velocity vector field reported in Fig.~\ref{fig2:Low1Curr}.
Indeed, the velocity field does not seem to be globally affected, although the localized multi-vacancy induces a very steep perturbation and the time averaged vector velocity calculated in the state localized at the lower edge perfectly bypasses the defect.
This behavior is distinctive of edge states, whereas the velocity vector field calculated in any other (not edge-localized) state is rather different, as shown in Fig.~\ref{fig3:bulkCurr}.

\begin{center}
\begin{figure}[H]
 \centering \includegraphics[width=9cm]{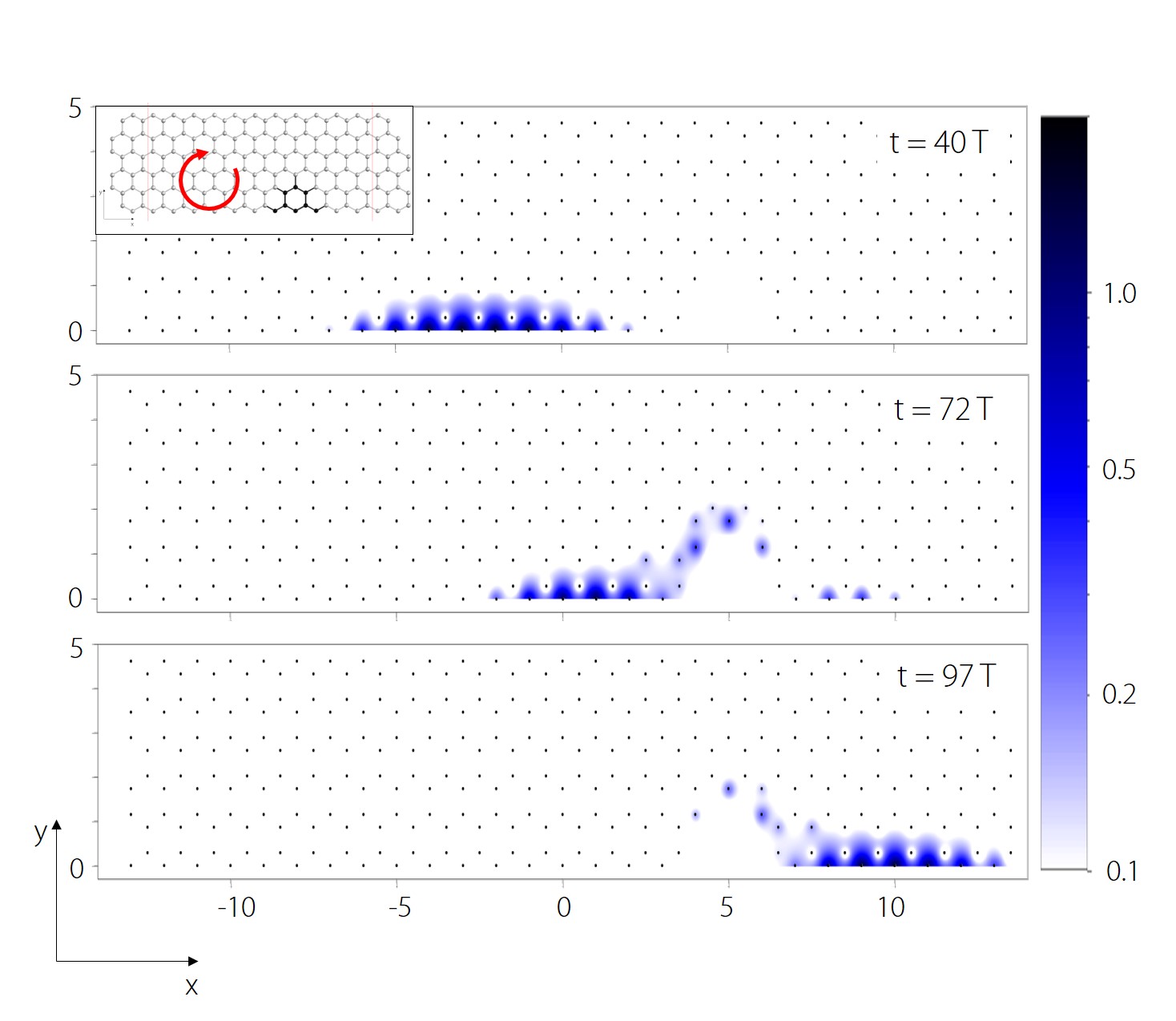}
 \caption{\label{fig4:Evo1} (color online) Snapshots at three different times (from top to bottom: $t = 40 \ T$, $t = 72 \ T$, $t = 97 \ T$) of the evolution of a state localized at the GNR lower edge in the presence of a multi-atom vacancy. The initial state is taken as a superposition of equally weighted edge states with $k \in [1.1,1.3] \ \pi$.
The color scale represents the probability density to find the electron on each site. }
\end{figure}
\end{center}

The real-time dynamics of the same edge state described in terms of hopping probability as given in Eq.~(\ref{probab}) is shown in Fig.~\ref{fig4:Evo1} and illustrates even more clearly how an electron injected in this state travels unaffected after circumventing the defect.

Topologically protected edge states have been proposed for the realization of quantum computing architectures based on the flying-qubit paradigm \cite{Stace2004,Giovannetti2008}, where electrons in chiral edge channels in the integer quantum Hall effect host and process quantum information.
In the above case, a magnetic field orthogonal to the plane of a quantum well, hosting a low-density electron gas, drives the 2D system in the quantum Hall regime.
A pattern of split gates creates the path of Hall edge states that can eventually lead to inter-channel interaction and coherent channel mixing \cite{Neder2007,Beggi2015}.
This  can be obtained with local magnetic fields \cite{Karmakar2011} and a single edge channel can be split by a proper quantum point contact \cite{Roddaro2003,Paradiso2011}.
Our proposal addresses the formation and control of edge states in GNRs with a high-frequency laser field.
Such edge states represent an alternative to the ones induced by a transverse magnetic field in a 2D semiconductor material through Landau quantization. 
Indeed, a remarkable difference between the two systems is the robustness of the former against local few-atoms defects. As it can be gathered from Fig.~\ref{fig4:Evo1}, the GNR edge state bypasses the perturbation preserving its original character, thus the localized excitation is neither scattered back nor excited to higher energy states. The reason can be traced to the presence of a single edge state in a given edge and to the large separation between the energy of our edge carrier and other available states with the same k vector (about $1$~eV in the system we simulated).
On the contrary, Hall edge states in a mesoscopic slab of a semiconductor, e.g. GaAs, are sensitive to sharp potential discontinuities, that may mix different edge channels \cite{PhysRevB.45.9059} (with different Landau index, though propagating in the same direction) whose energy difference is of the order of $\hbar\omega_c$, i.e. as small as $10$~meV for GaAs in a $5$~T magnetic field. 

\section{Dynamics of edge states in driven GNR with potential barrier}

We have shown that edge states of an ac-driven GNR are indeed robust against lattice defects. We now show how their real-space pattern can be controlled by adding a modulated electrostatic potential.
For simplicity, we consider a constant potential barrier crossing the whole width of the GNR and compute the edge-state dynamics in real time and real space for different heights of the barrier. Notice that the inclusion of the potential does not jeopardize the formation of edge states.

\begin{figure*}
 \centering \includegraphics[width=18cm]{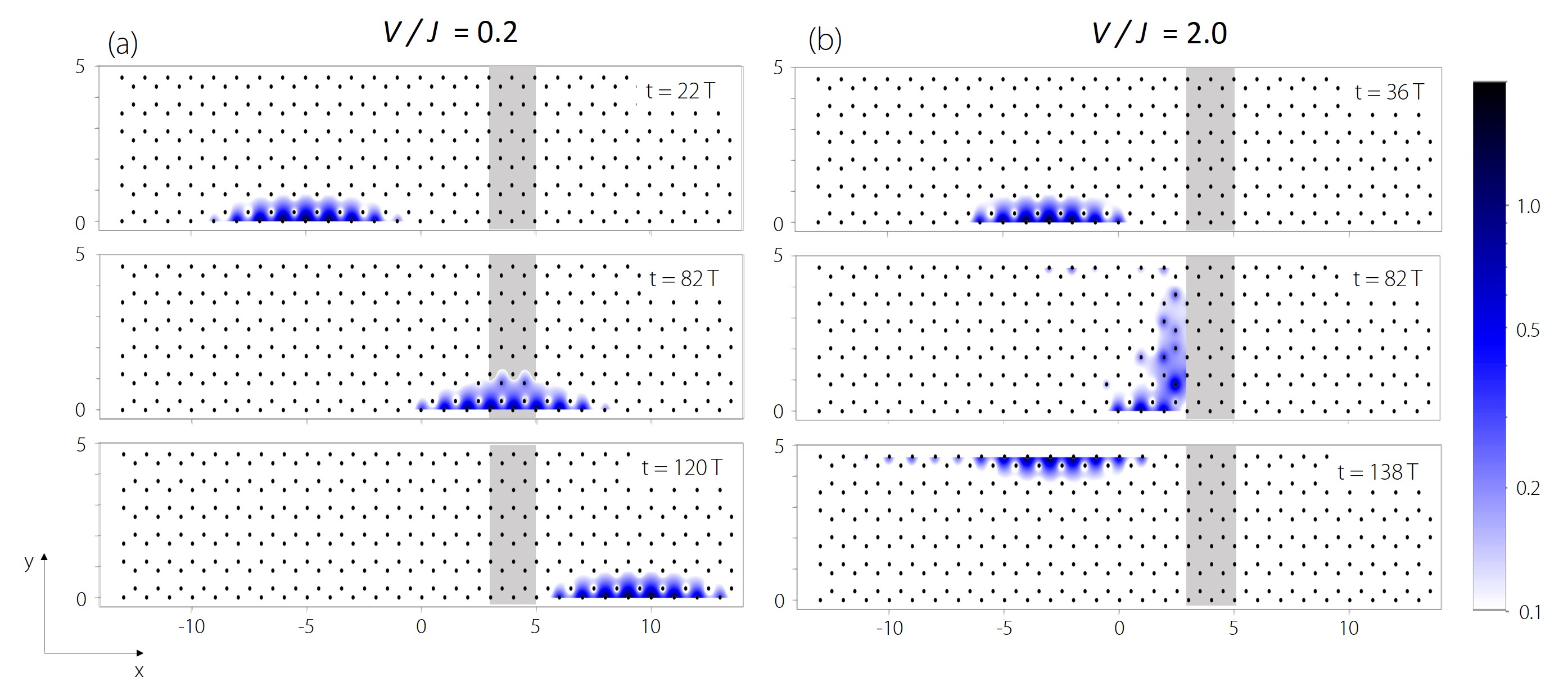}
 \caption{\label{fig5:Vstep1} (color online) Snapshots at different times of the dynamics of a right-mover lower-edge state in the presence of a potential barrier (dashed portion) $V(x)$ for $3 \leq x \leq 5$. The spatial coordinates are given in units of the lattice parameter $a$. (a) Dynamics with $V(x) = 2$ at times from top to bottom: $t = 36 \ T$, $t = 82 \ T$, $t = 138 \ T$. (b) Dynamics with $V(x) = 0.2$ at times from top to bottom: $t = 22 \ T$, $t = 82 \ T$, $t = 120 \ T$.}
\end{figure*}

\begin{center}
\begin{figure}[H]
 \centering \includegraphics[width=9cm]{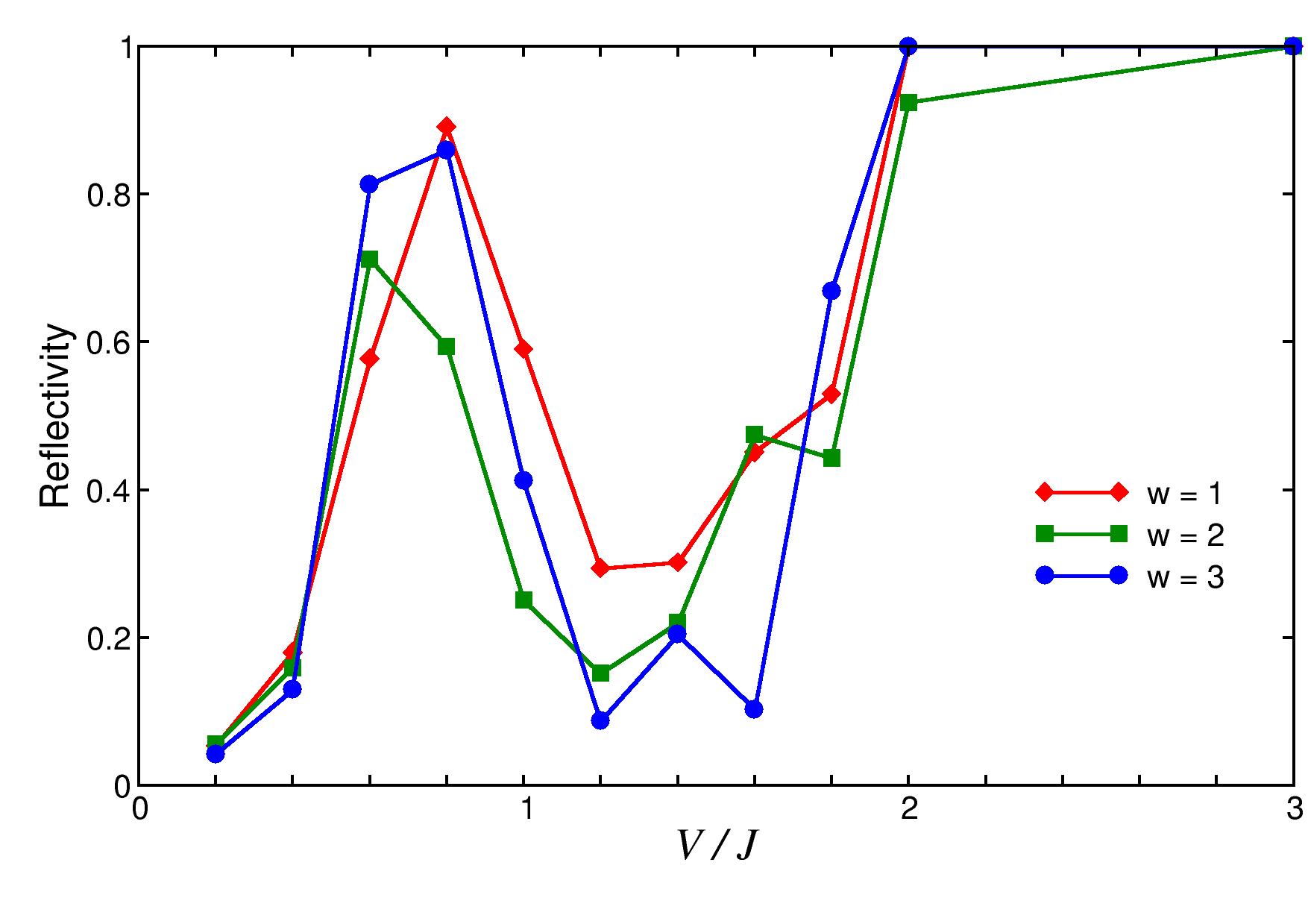}
 \caption{\label{fig6:Refl} (color online) Reflectivity as a function of the ratio between the height of the potential barrier $V$ and the hopping energy $J$ for different widths {\sffamily w} of the barrier.}
\end{figure}
\end{center}

As shown in  Fig.~\ref{fig5:Vstep1}, for a potential barrier with on-site energy $V = 0.2 \ J$, i.e. much lower than the inter-site hopping parameter $J$, the particle is perfectly transmitted, fully maintaining its edge character.  On the contrary, for a potential barrier substantially higher than the hopping term ($V = 2 J$) the electron localized on one edge is reflected onto the other GNR edge.
Although such a steep variation in the value of $V$ is beyond current technology, typically based on external gates leading to a smooth potential modulation, the tailoring of the edge channel path induced by space-dependent on-site energies is remarkable. The same effect can be obtained e.g. by removing a short part of the GNR, and it is also expected to be present for smoother barriers. Simulations of a realistic case of the latter type would require an exceedingly large supercell and are beyond the scope of the present work.
The possibility for a particle to be reflected from the lower to the upper edge following the barrier boundary is related again to the robustness of these topologically protected edge states.
In particular, a systematic calculation of reflectivity has been performed as a function of the height of the potential barrier, for three different widths (see Fig.~\ref{fig6:Refl}).
For all the widths considered, a peak in reflectivity around $V=0.8\ J$ appears, then a minimum occurs at around $V=1.2\ J$ before perfect reflection is reached for $V \geq 2\ J$.
This non-monotonic behavior is similar to the transmission of a one-dimensional barrier, with a notable difference.
On the one hand, in the trivial quantum mechanical problem of a square barrier in one dimension, resonance dips occur for  particle energies  slightly higher than the barrier ($E \gtrsim $ V), i.e. when the wave function has a plane-wave character in the barrier region.
On the other hand, in our system, we have a dip in the reflection coefficient when the potential barrier is higher than the hopping energy, i.e. when the wave function decays exponentially inside the barrier along $x$.
The observed drop in reflectivity associated with evanescent waves inside the barrier is evocative of a Klein tunneling mechanism \cite{Katsnelson2006,Allain2011}. 

All in all, a greater control shows up in the opportunity of having either transmitted particles on the same edge and/or reflected ones on the opposite edge of the nanoribbon. It is also important to notice that this behavior is persistent for longer potential steps or wider GNR's.

\section{Conclusion and outlook}

In conclusion, an efficient scheme to calculate the time evolution of Floquet states in real time and real space has been developed and applied to demonstrate the topological robustness of edge states in ac-driven graphene nanoribbons. Edge states evolve in time bypassing edge defects undisturbed.
Performing the evolution of these states in the presence of a potential barrier, we have shown that it is possible to control the edge carrier dynamics by tailoring the barrier height, in order to switch from a perfect transmission condition to a perfect reflection on the opposite edge of the GNR.
This is nontrivial and more intriguing compared to the analogous one-dimensional quantum mechanical problem.
This control mechanism can be exploited to realize topologically-protected flying qubits that might parallel similar proposals based on the quantum Hall effect.


%

\end{document}